\documentclass[trackchanges, twocolumn]{aastex7}

\begin{document}

\title{FastPM: Fast Pretrained Model for Cosmological Simulation-Based Inference}

\author[orcid=0000-0003-1585-997X]{Masaya Yamamoto}
\affiliation{Department of Astrophysical Sciences, Princeton University, Peyton Hall, Princeton, NJ 08544, USA}
\email[show]{masaya.yamamoto@princeton.edu}

\author[orcid=0000-0002-3568-3900]{Adrian E. Bayer}
\affiliation{The NSF AI Institute for Artificial Intelligence and Fundamental Interactions, Cambridge, MA 02139, USA}
\affiliation{Laboratory for Nuclear Science, Massachusetts Institute of Technology, Cambridge, MA 02139, USA}
\affiliation{Center for Astrophysics $|$ Harvard \& Smithsonian, 60 Garden Street, Cambridge, MA 02138, USA}
\affiliation{Perimeter Institute for Theoretical Physics, 31 Caroline Street North, Waterloo, Ontario N2L 2Y5, Canada}
\email[show]{abayer@mit.edu}

\begin{abstract}

Modern cosmology increasingly relies on large ensembles of expensive simulations to model the non-linear Universe and extract information from upcoming surveys. However, generating high-fidelity simulations—whether gravity-only N-body or full hydrodynamical—requires substantial computational resources. In this work, we demonstrate that transfer learning from fast low-fidelity simulations can dramatically reduce the cost of training machine learning models, in particular for simulation-based inference (SBI). We train a Fast Pretrained Model (FastPM) on a set of Fast Particle-Mesh (FastPM) simulations to infer $\Omega_m$ and $\sigma_8$ from the power spectrum, and finetune on a smaller set of high-fidelity CAMELS N-body and hydrodynamical simulations.
To reach a fixed posterior accuracy, the pretrained model requires only 5--15\% of the high-fidelity simulations compared to the number of simulations needed when training on CAMELS alone, reducing the total training simulation compute cost, including that of the pretraining simulations, by up to an order of magnitude in CPU hours. Remarkably, the gains persist for the total matter power spectrum of full hydrodynamical simulations: although FastPM contains no baryonic physics, pretraining on FastPM runs that together cost less than a single hydrodynamical simulation still reduces the total training simulation compute by a factor of $\approx$5. Moreover, pretraining on more accurate FastPM runs yields more accurate posteriors than training on CAMELS alone achieves with any number of simulations we consider. These results demonstrate that transfer learning from fast simulations is a simple and powerful route to high-fidelity SBI, and highlight the promise of simulation-based foundation models in cosmology.

\end{abstract}

\section{Introduction} 

Simulation-based inference (SBI) has rapidly gained traction in cosmological analyses as a powerful framework for extracting information beyond standard Gaussian summary statistics \citep[for a review, see][]{Thiele26}. By leveraging summary statistics that range from peak counts to voids, or by working at the field level, SBI enables the derivation of robust cosmological constraints from datasets where traditional analytical likelihoods are intractable \citep[e.g.,][]{Jeffrey21, Detecting, Porqueres23, Fluri22, Boruah22, Hahn23, Hahn23_2, Hahn24, Ho24, vonWietersheim25, Jeffrey25, NuCNN, arnab, SBI_LSST, huang2025cosmobenchmultiscalemultiviewmultitask, duivenvoorden2025robustcmbbmodeanalysis, bayer2025robust, Bayer:2026zcr, Williamson26, Thomsen26}.

This is typically achieved through algorithms such as neural posterior estimation (NPE, \citealt{Papamakarios16}), neural likelihood estimation (NLE; \citealt{papamakarios19a, lueckmann2019}), or neural ratio estimation (NRE; \citealt{hermans2020}). These models learn to approximate the likelihood or posterior directly by training on simulated datasets. 
Their performance depends on both the physical fidelity of the
simulations and the amount of training data. Simulator
inaccuracies can bias the inferred cosmological parameters
\citep[e.g.,][]{bayer2025robust}, while finite training sets
can limit posterior accuracy, precision, and calibration
\citep{Hermans21, Bairagi25, Homer25}. Accurate inference
therefore requires both sufficiently realistic forward models
and adequate training data.

Weak gravitational lensing is a particularly relevant
application, as it probes the projected total matter
distribution. Surveys now provide large galaxy shape
catalogues, such as DES Y6 \citep{Yamamoto25}, whose
cosmological information extends beyond two-point statistics.
Recent SBI analyses have extracted non-Gaussian information
from weak-lensing maps, improving cosmological constraints
beyond two-point analyses
\citep{Jeffrey25, Williamson26}.
Exploiting the information available on small scales requires
accurate modelling of nonlinear structure and baryonic
feedback, motivating methods that reduce the cost of training
on hydrodynamical simulations.

However, producing such comprehensive simulation suites is computationally prohibitive. Fully capturing the relevant physics—such as non-linear gravitational clustering and complex baryonic physics (hydrodynamics)—requires immense computational resources, making it practically infeasible to generate the sheer volume of high-fidelity simulations typically required to train deep neural networks from scratch.

To circumvent the prohibitive costs of large-scale, high-fidelity simulation grids, recent works have explored transfer learning strategies \citep{Thiele25, Saoulis25, Krishnaraj:2025fye, Krouglova25, Saoulis26}. The core concept is to pretrain a model using computationally cheap, low-fidelity simulations, and subsequently finetune the model using a scarce set of high-fidelity simulations that accurately capture the complex physics---or to train with multiple fidelities in one go. This approach has shown significant promise in accelerating cosmological parameter inference for both N-body power spectra \citep{Thiele25} and field-level inference from density maps in hydrodynamical simulations \citep{Saoulis25}, as well as in extending beyond the standard $\Lambda$CDM model \citep{Krishnaraj:2025fye}.

In this work, we train a Fast Pretrained Model (FastPM) with FastPM \citep{Feng2016, Bayer_2021_fastpm}. We demonstrate a framework to constrain cosmological parameters from the power spectrum, utilizing fast, approximate N-body simulations (FastPM) for pretraining, and high-fidelity simulations (CAMELS; \citealt{villaescusanavarro2020camels, CAMELS_public}) for finetuning. We develop transfer learning \textit{all the way} from cheap FastPM simulations to expensive hydrodynamic simulation---building on previous works transferring from FastPM to full N-body, or from full N-body to hydro. We also study the impact of the lowness of pretraining simulation fidelity by varying the FastPM number of simulation steps.

The paper is structured as follows. Section~\ref{sec:sims} describes the simulations we use in this work. Section~\ref{sec:model} describes the model architecture and training process. Section~\ref{sec:results} describes the results, and Section~\ref{sec:discussion} discusses the implications and paths for future work.

\section{Simulations}\label{sec:sims}
In this section, we describe the simulations used for transfer learning. We use FastPM simulations (described in \ref{subsec:fastpm}) to train a fast pretrained model to then finetune on CAMELS hydrodynamical simulation (described in \ref{subsec:camels}).
Table~\ref{tab:cost} summarizes the simulation configurations and computational cost of a single realization of each simulation type used in this work.
We now describe the simulations in more detail.

  \begin{deluxetable}{lccccc}
  \tablecaption{Cost of a single realization of each simulation used in this work.
  $N_{\rm p}$ is the number of particles, $N_{\rm mesh}$ the FastPM\ force mesh, with
  force-resolution factor $B=(N_{\rm mesh}/N_{\rm p})^{1/3}$ in parentheses, and $N_s$ the
  number of FastPM\ time steps (the high-fidelity codes use adaptive time stepping).
  At fixed mesh the FastPM\ cost scales approximately linearly with $N_s$
  ($\approx0.048$ CPU\,h per step).
  The FastPM runs with $N_s=200$ (20) are $\approx26\times$ ($\approx250\times$)
  cheaper than a CAMELS\ N-body run and $\approx320\times$ ($\approx3000\times$) cheaper
  than a CAMELS\ hydrodynamical run.
  \label{tab:cost}}
  \tablehead{
  \colhead{Simulation} & \colhead{Box [$h^{-1}$Mpc]} & \colhead{$N_{\rm p}$} &
  \colhead{$N_{\rm mesh}$ ($B$)} & \colhead{$N_s$} & \colhead{CPU h}
  }
  \startdata
  CAMELS\ N-body                & 25   & $256^3$        & --          & --  & 250 \\
  CAMELS\ hydro     & 25   & $2\times256^3$ & --          & --  & 3000 \\
  FastPM\      & 25   & $256^3$        & $512^3$ (2)      & 20       & 1.0 \\
  FastPM\     & 25   & $256^3$        & $512^3$ (2)      & 200      & 9.5 \\
  \enddata
  \end{deluxetable}

\begin{figure}
\centering
\includegraphics[width=\columnwidth]{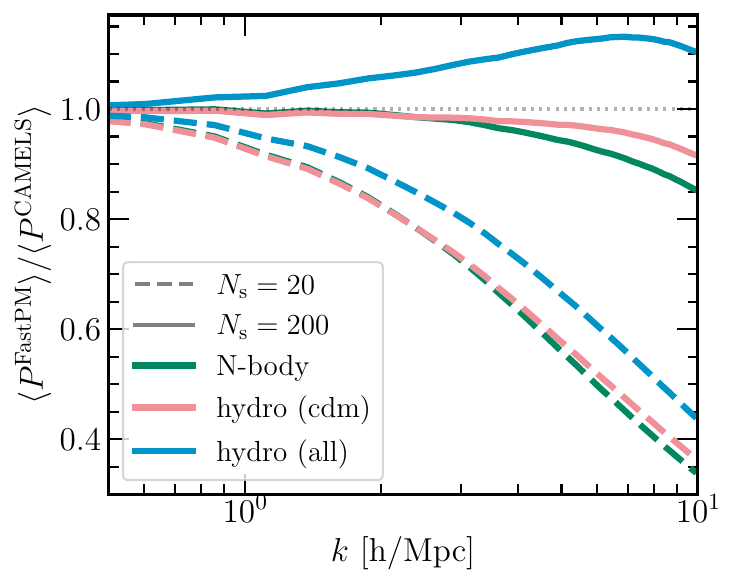}
    \caption{Ratio of the power spectrum from low-fidelity FastPM\ simulations (with number of steps $N_{\rm s}=20$ (dashed) and 200 (solid)) and high-fidelity CAMELS\ simulations, averaged over all simulations. We consider the N-body matter power, hydrodynamic CDM-only power, and hydrodynamic total matter power (CDM+baryons; all) from CAMELS.}
    \label{fig:ratio}
\end{figure}

\subsection{CAMELS}\label{subsec:camels}

The Cosmology and Astrophysics with Machine Learning Simulations (CAMELS) is a large suite of cosmological hydrodynamical simulations designed to enable and benchmark machine learning (ML) methods for cosmology and astrophysics \citep{ villaescusanavarro2020camels, CAMELS_public}. CAMELS\ provides simulations in terms of 9 hydrodynamic suites (e.g., IllustrisTNG; \citealt{illustristng}, SIMBA; \citealt{Dave2019_Simba}), and different sets of cosmological and subgrid parameters\footnote{\url{https://camels.readthedocs.io/en/latest/suites_sets.html}}. In our study, we utilize the Latin hypercube (LH) set in the IllustrisTNG suite with $256^3$ dark matter and $256^3$ gas particles and the volume of $(25\,h^{-1}\text{Mpc})^3$, which contains 1,000 simulations. We also use the matched N-body (dark matter only) simulations.

The LH simulations set in CAMELS\ varies both cosmological and astrophysical parameters. The cosmological parameters \{$\Omega_m$, $\sigma_8$\} are arranged so that $\Omega_m \in$ [0.10; 0.50], $\sigma_8 \in$ [0.60; 1.00] with fixed $\Omega_b = 0.049$, $h = 0.6711$ and $n_s = 0.9624$. Out of the 9 simulation suites in CAMELS, we focus on the IllustrisTNG\ suite at $z=0$ and leave a comparison with the other suites to future work. The astrophysical parameters in IllustrisTNG control stellar and AGN feedback processes, and vary in the range $A_{\rm SN1} \in$ [0.25; 4], $A_{\rm SN2} \in$ [0.5; 2], $A_{\rm AGN1} \in$ [0.25; 4], and $A_{\rm AGN2} \in$ [0.5; 2]. Each CAMELS\ IllustrisTNG\ hydrodynamical realization costs $\approx3{,}000$ CPU hours, and its matched N-body counterpart $\approx250$ CPU hours (Table~\ref{tab:cost}).

We consider three CAMELS outputs: pure N-body, cold dark matter (CDM) from the hydrodynamical run, and total matter from the hydrodynamical run (denoted ``all'', i.e.,~including baryons).
For the hydrodynamical datasets, we infer only
$\{\Omega_{\rm m},\sigma_8\}$, while the four feedback
parameters vary across the training simulations. The learned
cosmological posterior therefore implicitly marginalizes over
the baryonic parameters.

\subsection{FastPM}\label{subsec:fastpm}

FastPM \citep{Feng2016, Bayer_2021_fastpm} is a highly-scalable particle mesh (PM) N-body solver that implements a modified time-stepping scheme to enforce correct linear displacement (1LPT) evolution via adjusted kick and drift factors.

For our analysis, we generate $10^4$ FastPM\ realizations for each of $N_s=20$ and $N_s=200$, simulating the same volume and sampling the same cosmological prior as the CAMELS\ Latin Hypercube (LH) set (Sec.~\ref{subsec:camels}). We pretrain on each of these sets, as well as on a 2{,}000-realization subset of the $N_s=20$ set.

All FastPM\ runs adopt the fixed cosmological parameters of the CAMELS\ LH set, $\Omega_b=0.049$, $h=0.6711$, and $n_s=0.9624$, in a flat $\Lambda$CDM cosmology with $w=-1$ and massless neutrinos, with only $\Omega_m$ and $\sigma_8$ varied. For each cosmology, we compute the linear matter power spectrum at $z=0$, and generate initial particle displacements with 2LPT at $z_{\rm init}=9$. The particles are then evolved to $z=0$ in $N_s$ time steps spaced linearly in the scale factor $a$. Each realization uses an independent random seed; we do not match the phases of the CAMELS\ initial conditions.

For a given number of particles, the simulation accuracy in FastPM is primarily controlled by the number of time steps $N_s$. While increasing $N_s$ improves the transfer function and cross-correlation coefficient, it also requires more compute. We thus explicitly compare different $N_s$ to consider the trade-off between lower fidelity data vectors and number of simulations required for transfer, in terms of the total compute. 
We consider $N_s=20$ and 200.

Figure~\ref{fig:ratio} shows the ratio of the FastPM\ power spectrum to that of each CAMELS\ simulation type, averaged over all simulations. With only 20 steps, FastPM underpredicts the power on all scales. The total matter power agrees slightly better because of two competing effects: FastPM\ does not model baryonic suppression, so the loss in power from baryonic suppression in CAMELS\ partly compensates for the numerical suppression in FastPM. 
However, the numerical suppression depends on cosmology while the baryonic suppression depends on astrophysics, so this cancellation is not expected to help SBI of the cosmological parameters. 
With 200 steps, the agreement is much improved for the N-body and hydro CDM power, but FastPM\ now overpredicts the total matter power because the baryonic suppression in CAMELS\ is much larger than the remaining numerical suppression in FastPM.
Our CAMELS-matched FastPM\ runs require $\approx 9.5$ and $\approx 1$ CPU hour for $N_s=200$ and 20, respectively (Table~\ref{tab:cost}). This enables testing the accuracy vs computational cost tradeoff for transfer learning.

\section{Model architecture and training}\label{sec:model}
We utilize Neural Posterior Estimation (NPE, \citealt{Greenberg19, Lueckmann17, Papamakarios16}), which amortizes the inference process by training a conditional normalizing flow to approximate the posterior distribution $p(\theta | x)$, where $\theta$ represents cosmological parameters and $x$ denotes the power spectrum in our study. When training the normalizing flows, the power spectrum is additionally compressed into an embedding space using a fully-connected multi-layer perceptron (MLP) where the network can potentially find better summary statistics. These are trained together as one network. In the NPE framework, the density estimator $q_{\phi}(\theta | x)$ ($\phi$: neural network weight parameters) is learned by minimizing the expected Kullback-Leibler (KL) divergence $\mathbb{E}_{p(x)}[D_{\rm KL}(p(\theta | x) \,\|\, q_{\phi}(\theta | x))]$. The objective function is then expressed as
\begin{eqnarray}
    \mathcal{L}(\phi) &=& -\mathbb{E}_{p(\theta, x)}[\log q_{\phi}(\theta | x)]  \\
    &\approx& -\frac{1}{N_{\rm sim}} \sum_{i=1}^{N_{\rm sim}} \log q_{\phi}(\theta_i | x_i),
\end{eqnarray}
where $N_{\rm sim}$ is the total number of simulations used for training. We use a neural spline flow (NSF) as the conditional density estimator. When training from scratch, both for pretraining on the low-fidelity simulations and for the baseline trained only on high-fidelity data, we optimize the number of training epochs; the peak learning rate of the one-cycle schedule; the dimension of the learned summary, i.e. the output of the embedding network; the number of hidden layers and the width of the embedding MLP; the number of hidden features in each coupling layer of the NSF; the number of flow transforms; the number of residual blocks per transform; and the dropout probability. When finetuning, the architecture is inherited from the pretrained model, and only the number of epochs and the learning rate are optimized. These are optimized in 100 trials in the \textit{optuna} \citep{optuna} framework. 
Hyperparameters are selected using a common held-out set of
100 high-fidelity validation simulations, which are never used
for gradient updates. We use this fixed validation set to
compare the models after hyperparameter optimization; we do
not optimize the validation-set size. Reported high-fidelity
training counts exclude these validation simulations. The CPU-hour budgets include
the cost of generating the training simulations, including
pretraining simulations where applicable, but exclude the compute required to generate the
validation and test sets.

To study transfer learning, we have two models with which we perform cosmological parameter inference. One is our baseline model that is trained solely on the ``high-fidelity'' dataset. The ``high-fidelity'' dataset in our case is CAMELS. The other is our transfer learning approach, where we first pretrain on a ``low-fidelity'' dataset, and then finetune on a "high-fidelity'' dataset via weight initialization. The ``low-fidelity'' dataset in our case is FastPM. 
We finetune (or, for the baseline, train) the network on $N_{\rm sim}^{\rm high-fid.}\in\{10, 25, 50, 100, 200, 400, 600, 700\}$ CAMELS\ simulations, and hold out 100 simulations for validations and the same 200 simulations for testing. We use $k_{\rm max} = 5\,h\,\mathrm{Mpc}^{-1}$ throughout (we find similar performance up to $k_{\rm max} = 10\,h\,\mathrm{Mpc}^{-1}$, and did not test beyond). We quantify performance with the negative log test probability,
\begin{equation}
    {\rm NLTP} = -\frac{1}{N_{\rm test}} \sum_{j=1}^{N_{\rm test}} \log q_{\phi}(\theta_j | x_j),
\end{equation}
averaged over the $N_{\rm test}=200$ held-out simulations, with each parameter linearly rescaled to $[-1,1]$ over its prior range; a lower NLTP means the posterior assigns higher density to the true parameters.

All results below are reported as the mean over 3 independently initialized training runs, with error bars showing the standard deviation across the ensemble.

\section{Results}\label{sec:results}

\begin{figure*}
\centering
\includegraphics[width=0.9\textwidth]{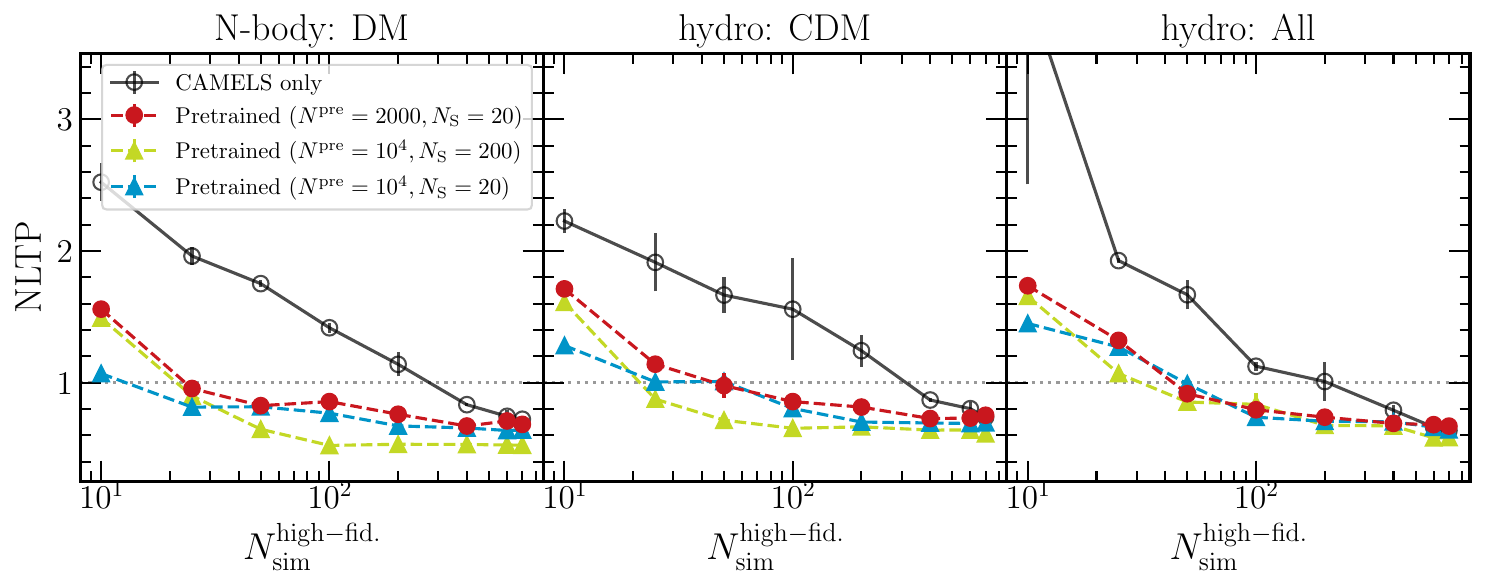}
\caption{The performance of the inference (NLTP) on the SBI models trained on only ``high-fidelity'' CAMELS\ simulations (solid black), and when pretrained on ``low-fidelity'' (FastPM) then finetuned on ``high-fidelity'' (CAMELS) simulations (dashed), as a function of the number of high-fidelity simulations ($N^{\rm high-fid.}_{\rm sim}$). The dotted line at NLTP=1 is the target NLTP at which we estimate the computational gain in the main text. From left to right column, the models are trained or finetuned using the power spectrum from N-body simulations, the CDM-only power spectrum in hydrodynamical simulations, and the total matter power spectrum in hydrodynamical simulations. Pretraining can be seen to require considerably fewer high-fidelity simulations.
}
\label{fig:fpm_camels}
\end{figure*} 

Figure~\ref{fig:fpm_camels} shows the NLTP as a function of the number of simulations for the various cases. 
We explicitly test how the fidelity of the pretraining set itself affects the result, varying both the number of FastPM\ realizations used in the training $N_{\rm sim}^{\rm pre}$ and the number of time steps $N_s$ used to generate them.
Comparing the solid black and colored dashed lines, pretraining clearly reduces the number of high-fidelity simulations required. Pretraining on a large set ($N_{\rm sim}^{\rm pre}=10^4$) of more accurate ($N_s=200$) FastPM\ runs gives the lowest NLTP at large $N_{\rm sim}^{\rm high-fid.}$, but even the cheapest configuration gives large savings: with $N_{\rm sim}^{\rm pre}=2{,}000$ and $N_s=20$, reaching NLTP$\,=1$ requires $\sim$90\%, 85\%, and $\sim$80\% fewer CAMELS\ simulations for the N-body, hydrodynamical CDM, and hydrodynamical total matter power spectra, respectively.
With the larger pretraining sets the reduction grows further, to as few as $\approx$5\%, 7\%, and 15\% of the CAMELS simulations required by the baseline.

\begin{figure*}
\centering
\includegraphics[width=0.9\textwidth]{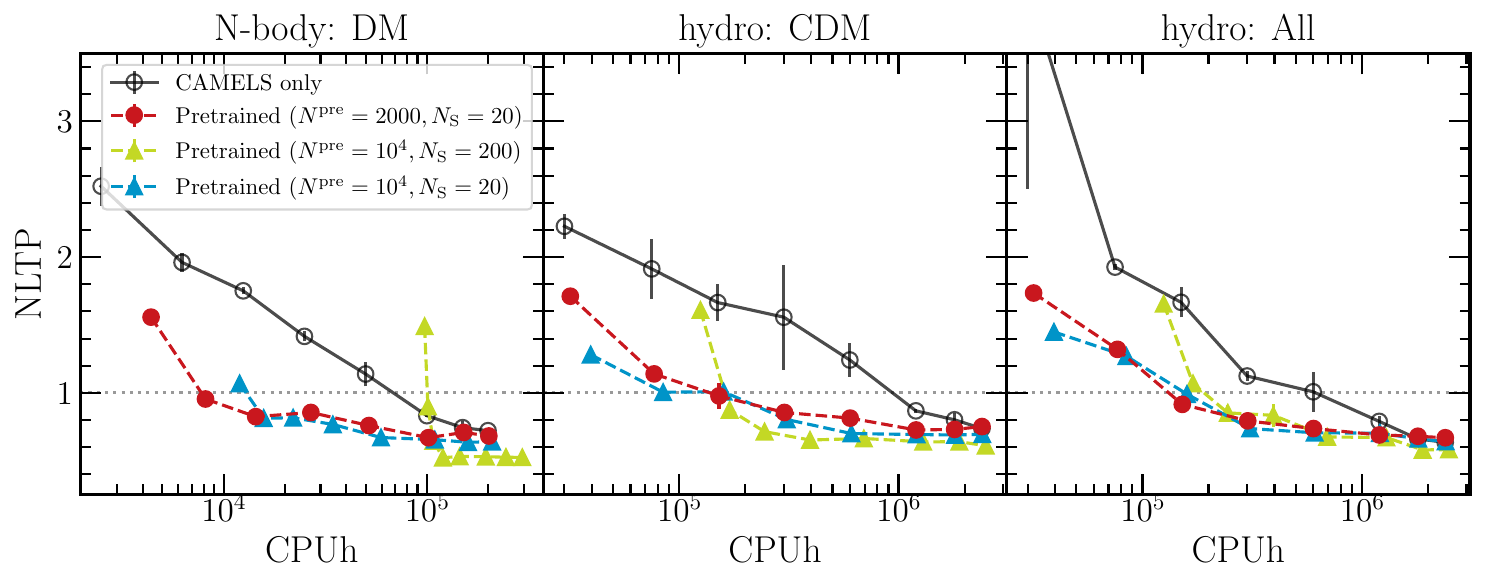}
\caption{
Like Figure \ref{fig:fpm_camels}, but as a function of CPU hours required to produce the simulations used for training (including for pretraining). Pretraining can be seen to require considerably fewer CPU hours: cheaper pretraining sets (red and blue) have the best performance when the simulation budget is low, with more expensive pretraining sets (green) overtaking for larger compute budgets.}
\label{fig:fpm_camels_ncpu}
\end{figure*}

Figure~\ref{fig:fpm_camels_ncpu} shows the NLTP as a function of the total number of CPU hours required to generate the simulations for both pretraining and finetuning. While a larger and more accurate set of FastPM simulations gives the best scaling with the number of finetuning simulations, it is also more expensive to run. Pretraining on $N_s=20$ runs saves compute in all three cases: with $N_{\rm sim}^{\rm pre}=2{,}000$, reaching NLTP$\,=1$ is $8.8\times$, $6.8\times$, and $4.7\times$ cheaper than training on CAMELS\ alone for the N-body, CDM-only, and total matter power spectra, respectively. 
Increasing the pretraining set size to $N_{\rm sim}^{\rm pre}=10^4$, still with step size $N_s=20$, raises the saving for the hydrodynamical CDM power spectrum to $\approx$10$\times$.
By contrast, the $N_s=200$ pretraining suite alone costs $\approx9.5\times10^4$ CPU hours, more than the $\approx7\times10^4$ CPU hours the baseline needs to reach NLTP$\,=1$ for the N-body power spectrum. This configuration however still saves compute for the hydrodynamical cases, where each CAMELS\ run costs 3{,}000 CPU hours, and it achieves the lowest NLTP at the largest training-set sizes explored here. Thus, when the simulation budget is limited, sacrificing fidelity in the pretraining simulations leads to significantly more efficient high-fidelity inference.

The extent of the performance gain directly correlates with the physical fidelity of FastPM relative to the target dataset. We observe the largest improvement for the N-body power spectrum, which aligns closely with the gravity-only physics captured by FastPM. Intermediate gains are seen for the CDM-only power spectrum from the hydrodynamical simulations, where the dark matter distribution is altered by baryonic back-reaction. Finally, the smallest improvement is reported for the total matter power spectrum; here, baryons contribute directly, and FastPM lacks the necessary baryonic physics. 
This pattern suggests that pretraining captures useful cosmological dependence from gravitational clustering, with finetuning adapting the model to baryonic effects.
Moreover, baryonic effects are degenerate with $\sigma_8$, which might lead to negative learning \citep{Krishnaraj:2025fye}, although we do not see this here.
For $N_{\rm sim}^{\rm high-fid.}\gtrsim400$, the baseline and the $N_s=20$ pretrained models reach a common NLTP of $\approx0.7$ in all three cases, so pretraining on approximate simulations accelerates convergence without degrading the accuracy ultimately reached. The model pretrained on $N_{\rm sim}^{\rm pre}=10^4$ runs with $N_s=200$ levels off lower, at NLTP$\,\approx0.5$ for the N-body power spectrum and $\approx0.6$ for the hydrodynamical cases. Since all models use the same input statistic and test set,
this difference indicates that the baseline plateau is not
simply a limit imposed by the information content of the
power spectrum. Higher-fidelity pretraining improves both
training efficiency and the best posterior accuracy achieved
within the simulation budgets explored here.

\begin{figure*}
\includegraphics[width=\textwidth]{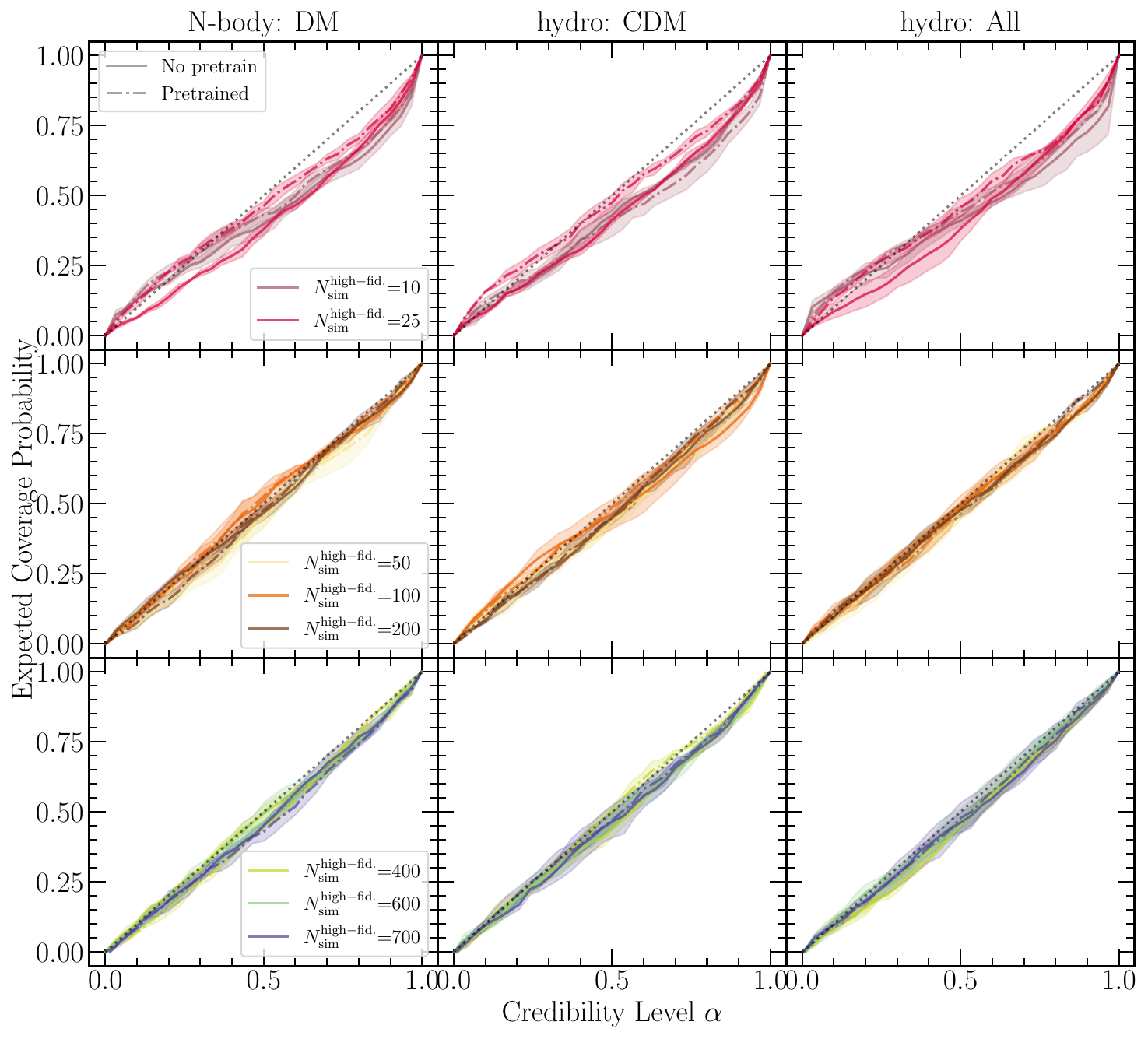}
\caption{Expected coverage probability computed using the Tests of Accuracy with Random Points (TARP) method for the posteriors generated for the $N_s=20$ and $N_{\rm sim}^{\rm pre}=2{,}000$ case. The panels are divided by the types of power spectrum: N-body dark matter (left column), cold dark matter only in hydrodynamical simulations (middle column), and total matter in hydrodynamical simulations (right column). Each row shows the posterior calibration for different numbers of ``high-fidelity'' simulations ($N^{\rm high-fid.}_{\text{sims}}$). Solid lines denote the baseline models trained exclusively on ``high-fidelity'' CAMELS\ simulations, while dash-dotted lines represent the multi-fidelity models finetuned from FastPM\ pretraining. The diagonal dotted line indicates a perfectly calibrated posterior where the expected coverage matches the credibility level $\alpha$. Shaded regions indicate $\pm1$ standard deviation in coverage
across the three independently initialized models.
}
\label{fig:tarp}
\end{figure*}

To show the calibration of our posteriors, we compute expected coverage probabilities using the Tests of Accuracy with Random Points (TARP) method \citep{2023PMLR..20219256L}. Figure~\ref{fig:tarp} shows the TARP results. For a perfectly calibrated posterior, the expected coverage probability equals the credibility level $\alpha$, tracing the diagonal; curves under the diagonal indicate over-confidence, in the sense that credible regions contain the true parameters less often
than their nominal level implies.
At the smallest training-set sizes
($N_{\rm sim}^{\rm high-fid.}=10,25$), both approaches exhibit
some overconfidence, with the pretrained models generally
showing better coverage. These departures generally
decrease as the training set grows. For
$N_{\rm sim}^{\rm high-fid.}\geq50$, both approaches closely
follow the diagonal in these tests, while pretraining
continues to improve NLTP (Figure~\ref{fig:fpm_camels}).

\section{Summary and Discussion}\label{sec:discussion}

We have shown that pretraining a neural posterior estimator on fast, approximate particle-mesh simulations and finetuning it on a small set of high-fidelity simulations substantially reduces the cost of cosmological parameter inference from the power spectrum. Using FastPM as the low-fidelity solver and CAMELS IllustrisTNG as the high-fidelity target, our main findings are:
\begin{enumerate}
    \item \emph{Cheap pretraining gives large savings.} Pretraining on just 2{,}000 20-step FastPM simulations, which together cost less than a single CAMELS hydrodynamical simulation, reduces the number of high-fidelity simulations needed to reach NLTP$\,=1$ by 80--90\% and the total training simulation CPU cost by 4.7--8.8$\times$ across the three different finetuning tasks. 
    \item \emph{Different pretraining fidelities offer complementary benefits.} When the simulation budget is limited, cheap 20-step runs give the largest compute savings. With a larger budget, pretraining on $10^4$ simulations with more accurate 200-step runs reaches a lower NLTP than the baseline attains with any number of high-fidelity CAMELS simulations we consider, improving the best posterior accuracy achieved within the simulation budgets explored here. More expensive pretraining may therefore be particularly attractive when a single pretrained model can be reused across multiple finetuning tasks, allowing its upfront cost to be amortized.
    \item \emph{Transfer works all the way to hydrodynamics.} The gains persist for the total matter power spectrum of full hydrodynamical simulations, even though FastPM contains no baryonic physics. The size of the gain follows how much of the relevant physics FastPM captures: it is largest for the N-body power spectrum, intermediate for the CDM power spectrum from the hydrodynamical simulations, and smallest for the total matter power spectrum in hydrodynamical simulations.
    
    \item \emph{Pretraining preserves or improves calibration.} Pretraining then finetuning on very few simulations gives comparable or slightly better empirical
    coverage than training on very few CAMELS simulations alone. Both are consistent with perfect calibration for $N_{\rm sim}^{\rm high-fidelity}\geq50$.
\end{enumerate}

Our results are obtained for the power spectrum, and it would be interesting to extend this approach to higher-order statistics and field-level inference. The dominant discrepancy between FastPM and exact N-body or hydrodynamical solutions lies in the transfer function rather than in the phases of the density field \citep{Sharma:2024kwj, Jia:2024wbr}, suggesting that the benefits of pretraining will extend
beyond the power spectrum. Moreover, transfer learning from N-body to hydrodynamical simulations has already been demonstrated for density maps \citep{Saoulis25}.

Several extensions could increase the gains further. The matched N-body and hydrodynamical simulations in CAMELS allow for transfer from FastPM to
hydrodynamics with a staged, or multi-fidelity, approach \citep[e.g.,][]{Thiele25, Saoulis25}. Additionally, applying the same framework to other CAMELS suites---or between CAMELS suites---such as SIMBA, whose feedback implementation differs substantially, will also be useful future work. Finally, pretraining on fast models that paint baryonic effects onto gravity-only simulations \citep[e.g.,][]{Schneider15, Dai21} is a promising route to extend these gains. 

Our results show that inexpensive, low-fidelity simulations can
provide an effective starting point for inference from much
more complex physical models. Gravity-only pretraining
transfers successfully to hydrodynamical matter power spectra,
with savings of up to an order of magnitude in the cost of
generating training simulations across the cases studied.
These gains are achieved using standard neural posterior
estimators, making the approach straightforward to incorporate
into existing SBI workflows. More broadly, this motivates
simulation-based foundation models pretrained on inexpensive
simulations and adapted with limited high-fidelity data to
new physical models, observables, and surveys, reducing the
need for large, bespoke training sets for each inference task.

\begin{acknowledgments}
We thank Leander Thiele, Shy Genel and Francisco Villaescusa-Navarro for useful discussions. The FastPM\ simulations were performed at facilities supported by the Scientific Computing Core at the Flatiron Institute, a division
of the Simons Foundation. Model training and inference was performed using Princeton University’s Research Computing resources. AEB is supported by the National Science Foundation under Cooperative Agreement PHY-2019786 (The NSF AI Institute for Artificial Intelligence and Fundamental Interactions). 
\end{acknowledgments}

\software{FastPM \citep{Feng2016},
          optuna \citep{optuna},
          tarp \citep{2023PMLR..20219256L}
          }

\bibliography{main}{}

@ARTICLE{Feng2016,
       author = {{Feng}, Yu and {Chu}, Man-Yat and {Seljak}, Uro{\v{s}} and
         {McDonald}, Patrick},
        title = "{FASTPM: a new scheme for fast simulations of dark matter and haloes}",
      journal = {\mnras},
         year = 2016,
        month = dec,
       volume = {463},
       number = {3},
        pages = {2273-2286},
          doi = {10.1093/mnras/stw2123},
archivePrefix = {arXiv},
       eprint = {1603.00476},
 primaryClass = {astro-ph.CO},
       adsurl = {https://ui.adsabs.harvard.edu/abs/2016MNRAS.463.2273F}
}

@article{Bayer_2021_fastpm,
	doi = {10.1088/1475-7516/2021/01/016},
	url = {https://doi.org/10.1088/1475-7516/2021/01/016},
	year = 2021,
	month = {jan},
	publisher = {{IOP} Publishing},
	volume = {2021},
	number = {01},
	pages = {016--016},
	author = {Adrian E. Bayer and Arka Banerjee and Yu Feng},
	title = {A fast particle-mesh simulation of non-linear cosmological structure formation with massive neutrinos},
	journal = {\jcap},
}

@ARTICLE{CAMELS_public,
       author = {{Villaescusa-Navarro}, Francisco and {Genel}, Shy and {Angl{\'e}s-Alc{\'a}zar}, Daniel and {Perez}, Lucia A. and {Villanueva-Domingo}, Pablo and {Wadekar}, Digvijay and {Shao}, Helen and {Mohammad}, Faizan G. and {Hassan}, Sultan and {Moser}, Emily and {Lau}, Erwin T. and {Machado Poletti Valle}, Luis Fernando and {Nicola}, Andrina and {Thiele}, Leander and {Jo}, Yongseok and {Philcox}, Oliver H.~E. and {Oppenheimer}, Benjamin D. and {Tillman}, Megan and {Hahn}, ChangHoon and {Kaushal}, Neerav and {Pisani}, Alice and {Gebhardt}, Matthew and {Delgado}, Ana Maria and {Caliendo}, Joyce and {Kreisch}, Christina and {Wong}, Kaze W.~K. and {Coulton}, William R. and {Eickenberg}, Michael and {Parimbelli}, Gabriele and {Ni}, Yueying and {Steinwandel}, Ulrich P. and {La Torre}, Valentina and {Dave}, Romeel and {Battaglia}, Nicholas and {Nagai}, Daisuke and {Spergel}, David N. and {Hernquist}, Lars and {Burkhart}, Blakesley and {Narayanan}, Desika and {Wandelt}, Benjamin and {Somerville}, Rachel S. and {Bryan}, Greg L. and {Viel}, Matteo and {Li}, Yin and {Irsic}, Vid and {Kraljic}, Katarina and {Vogelsberger}, Mark},
        title = "{The CAMELS project: public data release}",
      journal = {arXiv e-prints},
         year = 2022,
        month = jan,
          eid = {arXiv:2201.01300},
        pages = {arXiv:2201.01300},
archivePrefix = {arXiv},
       eprint = {2201.01300},
 primaryClass = {astro-ph.CO},
       adsurl = {https://ui.adsabs.harvard.edu/abs/2022arXiv220101300V}
}

@ARTICLE{villaescusanavarro2020camels,
       author = {{Villaescusa-Navarro}, Francisco and {Angl{\'e}s-Alc{\'a}zar}, Daniel and {Genel}, Shy and {Spergel}, David N. and {Somerville}, Rachel S. and {Dave}, Romeel and {Pillepich}, Annalisa and {Hernquist}, Lars and {Nelson}, Dylan and {Torrey}, Paul and {Narayanan}, Desika and {Li}, Yin and {Philcox}, Oliver and {La Torre}, Valentina and {Maria Delgado}, Ana and {Ho}, Shirley and {Hassan}, Sultan and {Burkhart}, Blakesley and {Wadekar}, Digvijay and {Battaglia}, Nicholas and {Contardo}, Gabriella and {Bryan}, Greg L.},
        title = "{The CAMELS Project: Cosmology and Astrophysics with Machine-learning Simulations}",
      journal = {\apj},
         year = 2021,
        month = jul,
       volume = {915},
       number = {1},
          eid = {71},
        pages = {71},
          doi = {10.3847/1538-4357/abf7ba},
archivePrefix = {arXiv},
       eprint = {2010.00619},
 primaryClass = {astro-ph.CO},
       adsurl = {https://ui.adsabs.harvard.edu/abs/2021ApJ...915...71V}
}

@ARTICLE{Dave2019_Simba,
       author = {{Dav{\'e}}, Romeel and {Angl{\'e}s-Alc{\'a}zar}, Daniel and
         {Narayanan}, Desika and {Li}, Qi and {Rafieferantsoa}, Mika H. and
         {Appleby}, Sarah},
        title = "{SIMBA: Cosmological simulations with black hole growth and feedback}",
      journal = {\mnras},
         year = "2019",
        month = "Jun",
       volume = {486},
       number = {2},
        pages = {2827-2849},
          doi = {10.1093/mnras/stz937},
archivePrefix = {arXiv},
       eprint = {1901.10203},
 primaryClass = {astro-ph.GA},
       adsurl = {https://ui.adsabs.harvard.edu/abs/2019MNRAS.486.2827D}
}

@inproceedings{Optuna,
    title={Optuna: A Next-generation Hyperparameter Optimization Framework},
    author={Akiba, Takuya and Sano, Shotaro and Yanase, Toshihiko and Ohta, Takeru and Koyama, Masanori},
    booktitle={Proceedings of the 25th {ACM} {SIGKDD} International Conference on Knowledge Discovery and Data Mining},
    year={2019}
}

@article{illustristng,
    author = "Nelson, Dylan and others",
    title = "{The IllustrisTNG simulations: public data release}",
    eprint = "1812.05609",
    archivePrefix = "arXiv",
    primaryClass = "astro-ph.GA",
    doi = "10.1186/s40668-019-0028-x",
    journal = "Comput. Astrophys. Cosmol.",
    volume = "6",
    number = "1",
    pages = "2",
    year = "2019"
}

@article{Jia:2024wbr,
    author = "Jia, He",
    title = "{Cosmological Analysis with Calibrated Neural Quantile Estimation and Approximate Simulators}",
    eprint = "2411.14748",
    archivePrefix = "arXiv",
    primaryClass = "astro-ph.CO",
    month = "11",
    year = "2024"
}

@misc{arnab,
      title={Interpreting Cosmological Information from Neural Networks in the Hydrodynamic Universe}, 
      author={Arnab Lahiry and Adrian E. Bayer and Francisco Villaescusa-Navarro},
      year={2025},
      eprint={2504.17839},
      archivePrefix={arXiv},
      primaryClass={astro-ph.CO},
      url={https://arxiv.org/abs/2504.17839}, 
}

@article{Sharma:2024kwj,
    author = "Sharma, Divij and Dai, Biwei and Villaescusa-Navarro, Francisco and Seljak, Uros",
    title = "{A field-level emulator for modelling baryonic effects across hydrodynamic simulations}",
    eprint = "2401.15891",
    archivePrefix = "arXiv",
    primaryClass = "astro-ph.CO",
    doi = "10.1093/mnras/staf355",
    journal = "Mon. Not. Roy. Astron. Soc.",
    volume = "538",
    number = "3",
    pages = "1415--1426",
    year = "2025"
}

@ARTICLE{2023PMLR..20219256L,
       author = {{Lemos}, Pablo and {Coogan}, Adam and {Hezaveh}, Yashar and {Perreault-Levasseur}, Laurence},
        title = "{Sampling-Based Accuracy Testing of Posterior Estimators for General Inference}",
      journal = {40th International Conference on Machine Learning},
         year = 2023,
        month = jan,
       volume = {202},
        pages = {19256-19273},
          doi = {10.48550/arXiv.2302.03026},
archivePrefix = {arXiv},
       eprint = {2302.03026},
 primaryClass = {stat.ML},
       adsurl = {https://ui.adsabs.harvard.edu/abs/2023PMLR..20219256L}
}

@ARTICLE{Thiele25,
       author = {{Thiele}, Leander and {Bayer}, Adrian E. and {Takeishi}, Naoya},
        title = "{Simulation-Efficient Cosmological Inference with Multi-Fidelity SBI}",
      journal = {arXiv e-prints},
         year = 2025,
        month = jul,
          eid = {arXiv:2507.00514},
        pages = {arXiv:2507.00514},
          doi = {10.48550/arXiv.2507.00514},
archivePrefix = {arXiv},
       eprint = {2507.00514},
 primaryClass = {astro-ph.CO},
       adsurl = {https://ui.adsabs.harvard.edu/abs/2025arXiv250700514T}
}

@ARTICLE{Saoulis25,
       author = {{Saoulis}, Alex A. and {Piras}, Davide and {Jeffrey}, Niall and {Mancini}, Alessio Spurio and {Ferreira}, Ana M.~G. and {Joachimi}, Benjamin},
        title = "{Transfer learning for multifidelity simulation-based inference in cosmology}",
      journal = {\mnras},
         year = 2025,
        month = sep,
       volume = {542},
       number = {4},
        pages = {3231-3245},
          doi = {10.1093/mnras/staf1436},
archivePrefix = {arXiv},
       eprint = {2505.21215},
 primaryClass = {astro-ph.CO},
       adsurl = {https://ui.adsabs.harvard.edu/abs/2025MNRAS.542.3231S}
}

@ARTICLE{Thiele26,
       author = {{Thiele}, Leander},
        title = "{Machine Learning Techniques for Astrophysics and Cosmology: Simulation-Based Inference}",
      journal = {arXiv e-prints},
         year = 2026,
        month = may,
          eid = {arXiv:2605.10719},
        pages = {arXiv:2605.10719},
          doi = {10.48550/arXiv.2605.10719},
archivePrefix = {arXiv},
       eprint = {2605.10719},
 primaryClass = {astro-ph.CO},
       adsurl = {https://ui.adsabs.harvard.edu/abs/2026arXiv260510719T}
}

@inproceedings{Papamakarios16,
author = {Papamakarios, George and Murray, Iain},
title = {Fast ε-free inference of simulation models with Bayesian conditional density estimation},
year = {2016},
isbn = {9781510838819},
publisher = {Curran Associates Inc.},
address = {Red Hook, NY, USA},
booktitle = {Proceedings of the 30th International Conference on Neural Information Processing Systems},
pages = {1036–1044},
numpages = {9},
location = {Barcelona, Spain},
series = {NIPS'16}
}

@ARTICLE{Jeffrey21,
       author = {{Jeffrey}, Niall and {Alsing}, Justin and {Lanusse}, Fran{\c{c}}ois},
        title = "{Likelihood-free inference with neural compression of DES SV weak lensing map statistics}",
      journal = {\mnras},
         year = 2021,
        month = feb,
       volume = {501},
       number = {1},
        pages = {954-969},
          doi = {10.1093/mnras/staa3594},
archivePrefix = {arXiv},
       eprint = {2009.08459},
 primaryClass = {astro-ph.CO},
       adsurl = {https://ui.adsabs.harvard.edu/abs/2021MNRAS.501..954J}
}

@ARTICLE{Porqueres23,
       author = {{Porqueres}, Natalia and {Heavens}, Alan and {Mortlock}, Daniel and {Lavaux}, Guilhem and {Makinen}, T. Lucas},
        title = "{Field-level inference of cosmic shear with intrinsic alignments and baryons}",
      journal = {arXiv e-prints},
         year = 2023,
        month = apr,
          eid = {arXiv:2304.04785},
        pages = {arXiv:2304.04785},
          doi = {10.48550/arXiv.2304.04785},
archivePrefix = {arXiv},
       eprint = {2304.04785},
 primaryClass = {astro-ph.CO},
       adsurl = {https://ui.adsabs.harvard.edu/abs/2023arXiv230404785P}
}

@ARTICLE{Boruah22,
       author = {{Boruah}, Supranta S. and {Rozo}, Eduardo and {Fiedorowicz}, Pier},
        title = "{Map-based cosmology inference with lognormal cosmic shear maps}",
      journal = {\mnras},
         year = 2022,
        month = nov,
       volume = {516},
       number = {3},
        pages = {4111-4122},
          doi = {10.1093/mnras/stac2508},
archivePrefix = {arXiv},
       eprint = {2204.13216},
 primaryClass = {astro-ph.CO},
       adsurl = {https://ui.adsabs.harvard.edu/abs/2022MNRAS.516.4111B}
}

@ARTICLE{Fluri22,
       author = {{Fluri}, Janis and {Kacprzak}, Tomasz and {Lucchi}, Aurelien and {Schneider}, Aurel and {Refregier}, Alexandre and {Hofmann}, Thomas},
        title = "{Full w CDM analysis of KiDS-1000 weak lensing maps using deep learning}",
      journal = {\prd},
         year = 2022,
        month = apr,
       volume = {105},
       number = {8},
          eid = {083518},
        pages = {083518},
          doi = {10.1103/PhysRevD.105.083518},
archivePrefix = {arXiv},
       eprint = {2201.07771},
 primaryClass = {astro-ph.CO},
       adsurl = {https://ui.adsabs.harvard.edu/abs/2022PhRvD.105h3518F}
}

@ARTICLE{vonWietersheim25,
       author = {{von Wietersheim-Kramsta}, Maximilian and {Lin}, Kiyam and {Tessore}, Nicolas and {Joachimi}, Benjamin and {Loureiro}, Arthur and {Reischke}, Robert and {Wright}, Angus H.},
        title = "{KiDS-SBI: Simulation-based inference analysis of KiDS-1000 cosmic shear}",
      journal = {\aap},
         year = 2025,
        month = feb,
       volume = {694},
          eid = {A223},
        pages = {A223},
          doi = {10.1051/0004-6361/202450487},
archivePrefix = {arXiv},
       eprint = {2404.15402},
 primaryClass = {astro-ph.CO},
       adsurl = {https://ui.adsabs.harvard.edu/abs/2025A&A...694A.223V}
}

@ARTICLE{Jeffrey25,
       author = {{Jeffrey}, N. and {Whiteway}, L. and {Gatti}, M. and {Williamson}, J. and {Alsing}, J. and {Porredon}, A. and {Prat}, J. and {Doux}, C. and {Jain}, B. and {Chang}, C. and {Cheng}, T.-Y. and {Kacprzak}, T. and {Lemos}, P. and {Alarcon}, A. and {Amon}, A. and {Bechtol}, K. and {Becker}, M.~R. and {Bernstein}, G.~M. and {Campos}, A. and {Carnero Rosell}, A. and {Chen}, R. and {Choi}, A. and {DeRose}, J. and {Drlica-Wagner}, A. and {Eckert}, K. and {Everett}, S. and {Fert{\'e}}, A. and {Gruen}, D. and {Gruendl}, R.~A. and {Herner}, K. and {Jarvis}, M. and {McCullough}, J. and {Myles}, J. and {Navarro-Alsina}, A. and {Pandey}, S. and {Raveri}, M. and {Rollins}, R.~P. and {Rykoff}, E.~S. and {S{\'a}nchez}, C. and {Secco}, L.~F. and {Sevilla-Noarbe}, I. and {Sheldon}, E. and {Shin}, T. and {Troxel}, M.~A. and {Tutusaus}, I. and {Varga}, T.~N. and {Yanny}, B. and {Yin}, B. and {Zuntz}, J. and {Aguena}, M. and {Allam}, S.~S. and {Alves}, O. and {Bacon}, D. and {Bocquet}, S. and {Brooks}, D. and {da Costa}, L.~N. and {Davis}, T.~M. and {De Vicente}, J. and {Desai}, S. and {Diehl}, H.~T. and {Ferrero}, I. and {Frieman}, J. and {Garc{\'\i}a-Bellido}, J. and {Gaztanaga}, E. and {Giannini}, G. and {Gutierrez}, G. and {Hinton}, S.~R. and {Hollowood}, D.~L. and {Honscheid}, K. and {Huterer}, D. and {James}, D.~J. and {Lahav}, O. and {Lee}, S. and {Marshall}, J.~L. and {Mena-Fern{\'a}ndez}, J. and {Miquel}, R. and {Pieres}, A. and {Plazas Malag{\'o}n}, A.~A. and {Roodman}, A. and {Sako}, M. and {Sanchez}, E. and {Sanchez Cid}, D. and {Smith}, M. and {Suchyta}, E. and {Swanson}, M.~E.~C. and {Tarle}, G. and {Tucker}, D.~L. and {Weaverdyck}, N. and {Weller}, J. and {Wiseman}, P. and {Yamamoto}, M.},
        title = "{Dark energy survey year 3 results: likelihood-free, simulation-based wCDM inference with neural compression of weak-lensing map statistics}",
      journal = {\mnras},
         year = 2025,
        month = jan,
       volume = {536},
       number = {2},
        pages = {1303-1322},
          doi = {10.1093/mnras/stae2629},
archivePrefix = {arXiv},
       eprint = {2403.02314},
 primaryClass = {astro-ph.CO},
       adsurl = {https://ui.adsabs.harvard.edu/abs/2025MNRAS.536.1303J}
}

@ARTICLE{Williamson26,
       author = {{Williamson}, J. and {Makinen}, T.~L. and {Porqueres}, N. and {Jeffrey}, N. and {Heavens}, A. and {Gatti}, M. and {Wandelt}, B.~D. and {Whiteway}, L. and {Prat}, J. and {Alarcon}, A. and {Amon}, A. and {Bechtol}, K. and {Becker}, M.~R. and {Bernstein}, G.~M. and {Campos}, A. and {Carnero Rosell}, A. and {Chen}, R. and {Choi}, A. and {DeRose}, J. and {Doux}, C. and {Drlica-Wagner}, A. and {Eckert}, K. and {Everett}, S. and {Fert{\'e}}, A. and {Gong}, Z. and {Gruen}, D. and {Gruendl}, R.~A. and {Herner}, K. and {Jarvis}, M. and {Kacprzak}, T. and {Lahav}, O. and {McCullough}, J. and {Myles}, J. and {Navarro-Alsina}, A. and {Pandey}, S. and {Raveri}, M. and {Rollins}, R.~P. and {Rykoff}, E.~S. and {S{\'a}nchez}, C. and {Secco}, L.~F. and {Sevilla-Noarbe}, I. and {Sheldon}, E. and {Shin}, T. and {Thomsen}, A. and {Troxel}, M.~A. and {Tutusaus}, I. and {Varga}, T.~N. and {Yanny}, B. and {Yin}, B. and {Zuntz}, J. and {Abbott}, T.~M.~C. and {Aguena}, M. and {Andrade-Oliveira}, F. and {Brooks}, D. and {Camilleri}, R. and {Carretero}, J. and {Cawthon}, R. and {Crocce}, M. and {da Costa}, L.~N. and {Davis}, T.~M. and {De Vicente}, J. and {Desai}, S. and {Diehl}, H.~T. and {Flaugher}, B. and {Frieman}, J. and {Garc{\'\i}a-Bellido}, J. and {Gutierrez}, G. and {Hinton}, S.~R. and {Hollowood}, D.~L. and {Kuehn}, K. and {Marshall}, J.~L. and {Mena-Fern{\'a}ndez}, J. and {Miquel}, R. and {Mohr}, J.~J. and {Muir}, J. and {Plazas Malag{\'o}n}, A.~A. and {Porredon}, A. and {Roodman}, A. and {Sanchez}, E. and {Sanchez Cid}, D. and {Suchyta}, E. and {Swanson}, M.~E.~C. and {To}, C. and {Tucker}, D.~L. and {Vikram}, V. and {Walker}, A.~R. and {Weaverdyck}, N. and {Weller}, J.},
        title = "{Dark Energy Survey Year 3 results: optimized $w$CDM simulation-based inference with weak lensing map-level hybrid statistics}",
      journal = {arXiv e-prints},
         year = 2026,
        month = jun,
          eid = {arXiv:2606.11309},
        pages = {arXiv:2606.11309},
          doi = {10.48550/arXiv.2606.11309},
archivePrefix = {arXiv},
       eprint = {2606.11309},
 primaryClass = {astro-ph.CO},
       adsurl = {https://ui.adsabs.harvard.edu/abs/2026arXiv260611309W}
}

@ARTICLE{Greenberg19,
       author = {{Greenberg}, David S. and {Nonnenmacher}, Marcel and {Macke}, Jakob H.},
        title = "{Automatic Posterior Transformation for Likelihood-Free Inference}",
      journal = {arXiv e-prints},
         year = 2019,
        month = may,
          eid = {arXiv:1905.07488},
        pages = {arXiv:1905.07488},
          doi = {10.48550/arXiv.1905.07488},
archivePrefix = {arXiv},
       eprint = {1905.07488},
 primaryClass = {cs.LG},
       adsurl = {https://ui.adsabs.harvard.edu/abs/2019arXiv190507488G}
}

@ARTICLE{Lueckmann17,
       author = {{Lueckmann}, Jan-Matthis and {Goncalves}, Pedro J. and {Bassetto}, Giacomo and {{\"O}cal}, Kaan and {Nonnenmacher}, Marcel and {Macke}, Jakob H.},
        title = "{Flexible statistical inference for mechanistic models of neural dynamics}",
      journal = {arXiv e-prints},
         year = 2017,
        month = nov,
          eid = {arXiv:1711.01861},
        pages = {arXiv:1711.01861},
          doi = {10.48550/arXiv.1711.01861},
archivePrefix = {arXiv},
       eprint = {1711.01861},
 primaryClass = {stat.ML},
       adsurl = {https://ui.adsabs.harvard.edu/abs/2017arXiv171101861L}
}

@article{Krishnaraj:2025fye,
    author = "Krishnaraj, Veena and Bayer, Adrian E. and Jespersen, Christian Kragh and Melchior, Peter",
    title = "{Transfer learning beyond the standard model}",
    eprint = "2510.19168",
    archivePrefix = "arXiv",
    primaryClass = "astro-ph.CO",
    doi = "10.1088/1475-7516/2026/06/026",
    journal = "JCAP",
    volume = "06",
    pages = "026",
    year = "2026"
}

@InProceedings{papamakarios19a,
  title = 	 {Sequential Neural Likelihood: Fast Likelihood-free Inference with Autoregressive Flows},
  author =       {Papamakarios, George and Sterratt, David and Murray, Iain},
  booktitle = 	 {Proceedings of the Twenty-Second International Conference on Artificial Intelligence and Statistics},
  pages = 	 {837--848},
  year = 	 {2019},
  editor = 	 {Chaudhuri, Kamalika and Sugiyama, Masashi},
  volume = 	 {89},
  series = 	 {Proceedings of Machine Learning Research},
  month = 	 {16--18 Apr},
  publisher =    {PMLR},
  url = 	 {https://proceedings.mlr.press/v89/papamakarios19a.html}
}

@misc{lueckmann2019,
      title={Likelihood-free inference with emulator networks}, 
      author={Jan-Matthis Lueckmann and Giacomo Bassetto and Theofanis Karaletsos and Jakob H. Macke},
      year={2019},
      eprint={1805.09294},
      archivePrefix={arXiv},
      primaryClass={stat.ML},
      url={https://arxiv.org/abs/1805.09294}, 
}

@misc{hermans2020,
      title={Likelihood-free MCMC with Amortized Approximate Ratio Estimators}, 
      author={Joeri Hermans and Volodimir Begy and Gilles Louppe},
      year={2020},
      eprint={1903.04057},
      archivePrefix={arXiv},
      primaryClass={stat.ML},
      url={https://arxiv.org/abs/1903.04057}, 
}

@ARTICLE{Schneider15,
       author = {{Schneider}, Aurel and {Teyssier}, Romain},
        title = "{A new method to quantify the effects of baryons on the matter power spectrum}",
      journal = {\jcap},
         year = 2015,
       volume = {2015},
       number = {12},
          eid = {049},
          doi = {10.1088/1475-7516/2015/12/049},
archivePrefix = {arXiv},
       eprint = {1510.06034},
 primaryClass = {astro-ph.CO}
}

@ARTICLE{Dai21,
       author = {{Dai}, Biwei and {Seljak}, Uro{\v{s}}},
        title = "{Learning effective physical laws for generating cosmological hydrodynamics with Lagrangian deep learning}",
      journal = {Proceedings of the National Academy of Sciences},
         year = 2021,
       volume = {118},
       number = {16},
          eid = {e2020324118},
          doi = {10.1073/pnas.2020324118},
archivePrefix = {arXiv},
       eprint = {2010.02926},
 primaryClass = {astro-ph.CO}
}

@ARTICLE{Detecting,
       author = {{Bayer}, Adrian E. and {Villaescusa-Navarro}, Francisco and {Massara}, Elena and {Liu}, Jia and {Spergel}, David N. and {Verde}, Licia and {Wandelt}, Benjamin D. and {Viel}, Matteo and {Ho}, Shirley},
        title = "{Detecting Neutrino Mass by Combining Matter Clustering, Halos, and Voids}",
      journal = {\apj},
         year = 2021,
        month = sep,
       volume = {919},
       number = {1},
          eid = {24},
        pages = {24},
          doi = {10.3847/1538-4357/ac0e91},
archivePrefix = {arXiv},
       eprint = {2102.05049},
 primaryClass = {astro-ph.CO},
       adsurl = {https://ui.adsabs.harvard.edu/abs/2021ApJ...919...24B}
}

@article{SBI_LSST,
      title={Simulation-Based Inference Benchmark for LSST Weak Lensing Cosmology}, 
      author={Justine Zeghal and Denise Lanzieri and François Lanusse and Alexandre Boucaud and Gilles Louppe and Eric Aubourg and Adrian E. Bayer and {The LSST Dark Energy Science Collaboration}},
      year={2024},
      eprint={2409.17975},
      archivePrefix={arXiv},
      primaryClass={astro-ph.CO},
      url={https://arxiv.org/abs/2409.17975}, 
}

@article{NuCNN,
    author = "Golshan, Malika and Bayer, Adrian E.",
    title = "{Massive \ensuremath{\nu}s through the CNN lens: interpreting the field-level neutrino mass information in weak lensing}",
    eprint = "2410.00914",
    archivePrefix = "arXiv",
    primaryClass = "astro-ph.CO",
    doi = "10.1088/1475-7516/2025/05/024",
    journal = "JCAP",
    volume = "05",
    pages = "024",
    year = "2025"
}

@article{bayer2025robust,
    author = "Bayer, Adrian E. and Villaescusa-Navarro, Francisco and Sharief, Sammy and Teyssier, Romain and Garrison, Lehman H. and Perreault-Levasseur, Laurence and Bryan, Greg L. and Gatti, Marco and Visbal, Eli",
    title = "{Field-level Comparison and Robustness Analysis of Cosmological N-body Simulations}",
    eprint = "2505.13620",
    archivePrefix = "arXiv",
    primaryClass = "astro-ph.CO",
    doi = "10.3847/1538-4357/adef4e",
    journal = "ApJ",
    volume = "989",
    number = "2",
    pages = "207",
    year = "2025"
}

@article{huang2025cosmobenchmultiscalemultiviewmultitask,
      title={CosmoBench: A Multiscale, Multiview, Multitask Cosmology Benchmark for Geometric Deep Learning}, 
      author={Ningyuan Huang and Richard Stiskalek and Jun-Young Lee and Adrian E. Bayer and Charles C. Margossian and Christian Kragh Jespersen and Lucia A. Perez and Lawrence K. Saul and Francisco Villaescusa-Navarro},
      year={2025},
      eprint={2507.03707},
      archivePrefix={arXiv},
      primaryClass={cs.LG},
      url={https://arxiv.org/abs/2507.03707}, 
}

@article{duivenvoorden2025robustcmbbmodeanalysis,
      title={Robust CMB B-mode analysis with Needlet-ILC and simulation-based inference}, 
      author={Adriaan J. Duivenvoorden and Kristen Surrao and Adrian E. Bayer and Alexandre E. Adler and Nadia Dachlythra and Susanna Azzoni and J. Colin Hill},
      month={12},
      year={2025},
      eprint={2512.16869},
      archivePrefix={arXiv},
      primaryClass={astro-ph.CO},
}

@article{Bayer:2026zcr,
    author = "Bayer, Adrian E. and Parker, Liam and Valcin, David and Chen, Shi-Fan and Modi, Chirag and Seljak, Uros",
    title = "{Field-Level Inference from Galaxies: BAO Reconstruction}",
    eprint = "2603.15732",
    archivePrefix = "arXiv",
    primaryClass = "astro-ph.CO",
    month = "3",
    year = "2026"
}

@ARTICLE{Hermans21,
       author = {{Hermans}, Joeri and {Delaunoy}, Arnaud and {Rozet}, Fran{\c{c}}ois and {Wehenkel}, Antoine and {Begy}, Volodimir and {Louppe}, Gilles},
        title = "{A Trust Crisis In Simulation-Based Inference? Your Posterior Approximations Can Be Unfaithful}",
      journal = {arXiv e-prints},
         year = 2021,
        month = oct,
          eid = {arXiv:2110.06581},
        pages = {arXiv:2110.06581},
          doi = {10.48550/arXiv.2110.06581},
archivePrefix = {arXiv},
       eprint = {2110.06581},
 primaryClass = {stat.ML},
       adsurl = {https://ui.adsabs.harvard.edu/abs/2021arXiv211006581H}
}

@ARTICLE{Bairagi25,
       author = {{Bairagi}, Anirban and {Wandelt}, Benjamin and {Villaescusa-Navarro}, Francisco},
        title = "{The BIG SOBOL SEQUENCE: How many simulations do we need for simulation-based inference in cosmology?}",
      journal = {\aap},
         year = 2025,
        month = nov,
       volume = {703},
          eid = {A301},
        pages = {A301},
          doi = {10.1051/0004-6361/202554602},
archivePrefix = {arXiv},
       eprint = {2503.13755},
 primaryClass = {astro-ph.CO},
       adsurl = {https://ui.adsabs.harvard.edu/abs/2025A&A...703A.301B}
}

@ARTICLE{Homer25,
       author = {{Homer}, J. and {Friedrich}, O. and {Gruen}, D.},
        title = "{Simulation-based inference has its own Dodelson─Schneider effect (but it knows that it does)}",
      journal = {\aap},
         year = 2025,
        month = jul,
       volume = {699},
          eid = {A213},
        pages = {A213},
          doi = {10.1051/0004-6361/202453339},
archivePrefix = {arXiv},
       eprint = {2412.02311},
 primaryClass = {astro-ph.CO},
       adsurl = {https://ui.adsabs.harvard.edu/abs/2025A&A...699A.213H}
}

@ARTICLE{Krouglova25,
       author = {{Krouglova}, Anastasia N. and {Johnson}, Hayden R. and {Confavreux}, Basile and {Deistler}, Michael and {Gon{\c{c}}alves}, Pedro J.},
        title = "{Multifidelity Simulation-based Inference for Computationally Expensive Simulators}",
      journal = {arXiv e-prints},
         year = 2025,
        month = feb,
          eid = {arXiv:2502.08416},
        pages = {arXiv:2502.08416},
          doi = {10.48550/arXiv.2502.08416},
archivePrefix = {arXiv},
       eprint = {2502.08416},
 primaryClass = {stat.ML},
       adsurl = {https://ui.adsabs.harvard.edu/abs/2025arXiv250208416K}
}

@ARTICLE{Saoulis26,
       author = {{Saoulis}, Alex A. and {Lin}, Kiyam and {Jeffrey}, Niall and {von Wietersheim-Kramsta}, Maximilian and {Piras}, Davide and {Mancini}, Alessio Spurio and {Ferreira}, Ana M.~G. and {Joachimi}, Benjamin},
        title = "{Field-level weak lensing cosmology with 60 simulations using multifidelity simulation-based inference}",
      journal = {\mnras},
         year = 2026,
        month = sep,
          doi = {10.1093/mnras/stag1702},
archivePrefix = {arXiv},
       eprint = {2606.23346},
 primaryClass = {astro-ph.CO},
       adsurl = {https://ui.adsabs.harvard.edu/abs/2026MNRAS.tmp.1578S}
}

@ARTICLE{Ho24,
       author = {{Ho}, Matthew and {Bartlett}, Deaglan J. and {Chartier}, Nicolas and {Cuesta-Lazaro}, Carolina and {Ding}, Simon and {Lapel}, Axel and {Lemos}, Pablo and {Lovell}, Christopher C. and {Makinen}, T. Lucas and {Modi}, Chirag and {Pandya}, Viraj and {Pandey}, Shivam and {Perez}, Lucia A. and {Wandelt}, Benjamin and {Bryan}, Greg L.},
        title = "{LtU-ILI: An All-in-One Framework for Implicit Inference in Astrophysics and Cosmology}",
      journal = {The Open Journal of Astrophysics},
         year = 2024,
        month = jul,
       volume = {7},
          eid = {54},
        pages = {54},
          doi = {10.33232/001c.120559},
archivePrefix = {arXiv},
       eprint = {2402.05137},
 primaryClass = {astro-ph.IM},
       adsurl = {https://ui.adsabs.harvard.edu/abs/2024OJAp....7E..54H}
}

@ARTICLE{Yamamoto25,
       author = {{Yamamoto}, M. and {Becker}, M.~R. and {Sheldon}, E. and {Jarvis}, M. and {Gruendl}, R.~A. and {Menanteau}, F. and {Rykoff}, E.~S. and {Mau}, S. and {Schutt}, T. and {Gatti}, M. and {Troxel}, M.~A. and {Amon}, A. and {Anbajagane}, D. and {Bernstein}, G.~M. and {Gruen}, D. and {Huff}, E.~M. and {Tabbutt}, M. and {Tong}, A. and {Yanny}, B. and {Abbott}, T.~M.~C. and {Aguena}, M. and {Alarcon}, A. and {Andrade-Oliveira}, F. and {Bechtol}, K. and {Blazek}, J. and {Brooks}, D. and {Rosell}, A. Carnero and {Carretero}, J. and {Chang}, C. and {Choi}, A. and {Costanzi}, M. and {Crocce}, M. and {da Costa}, L.~N. and {Davis}, T.~M. and {De Vicente}, J. and {Desai}, S. and {Diehl}, H.~T. and {Dodelson}, S. and {Doel}, P. and {Doux}, C. and {Drlica-Wagner}, A. and {Fert{\'e}}, A. and {Flaugher}, B. and {Fosalba}, P. and {Frieman}, J. and {Garc{\'\i}a-Bellido}, J. and {Gaztanaga}, E. and {Giannini}, G. and {Gutierrez}, G. and {Hartley}, W.~G. and {Herner}, K. and {Hinton}, S.~R. and {Hollowood}, D.~L. and {Honscheid}, K. and {Huterer}, D. and {Krause}, E. and {Kuehn}, K. and {Lahav}, O. and {Lima}, M. and {Marshall}, J.~L. and {Mena-Fern{\'a}ndez}, J. and {Miquel}, R. and {Mohr}, J.~J. and {Muir}, J. and {Myles}, J. and {Ogando}, R.~L.~C. and {Pieres}, A. and {Malag{\'o}n}, A.~A. Plazas and {Porredon}, A. and {Prat}, J. and {Raveri}, M. and {Rodriguez-Monroy}, M. and {Roodman}, A. and {Samuroff}, S. and {Sanchez}, E. and {Sanchez Cid}, D. and {Scarpine}, V. and {Sevilla-Noarbe}, I. and {Smith}, M. and {Soares-Santos}, M. and {Suchyta}, E. and {Tarle}, G. and {Vikram}, V. and {Weaverdyck}, N. and {Wiseman}, P. and {Zhang}, Y.},
        title = "{Dark Energy Survey Year 6 results: cell-based coadds and METADETECTION weak lensing shape catalogue}",
      journal = {\mnras},
         year = 2025,
        month = nov,
       volume = {543},
       number = {4},
        pages = {4156-4186},
          doi = {10.1093/mnras/staf1661},
archivePrefix = {arXiv},
       eprint = {2501.05665},
 primaryClass = {astro-ph.CO},
       adsurl = {https://ui.adsabs.harvard.edu/abs/2025MNRAS.543.4156Y}
}

@ARTICLE{Hahn24,
       author = {{Hahn}, ChangHoon and {Lemos}, Pablo and {Parker}, Liam and {R{\'e}galdo-Saint Blancard}, Bruno and {Eickenberg}, Michael and {Ho}, Shirley and {Hou}, Jiamin and {Massara}, Elena and {Modi}, Chirag and {Moradinezhad Dizgah}, Azadeh and {Spergel}, David},
        title = "{Cosmological constraints from non-Gaussian and nonlinear galaxy clustering using the SIMBIG inference framework}",
      journal = {Nature Astronomy},
         year = 2024,
        month = nov,
       volume = {8},
        pages = {1457-1467},
          doi = {10.1038/s41550-024-02344-2},
       adsurl = {https://ui.adsabs.harvard.edu/abs/2024NatAs...8.1457H}
}

@ARTICLE{Hahn23,
       author = {{Hahn}, ChangHoon and {Eickenberg}, Michael and {Ho}, Shirley and {Hou}, Jiamin and {Lemos}, Pablo and {Massara}, Elena and {Modi}, Chirag and {Moradinezhad Dizgah}, Azadeh and {R{\'e}galdo-Saint Blancard}, Bruno and {Abidi}, Muntazir M.},
        title = "{SIMBIG: mock challenge for a forward modeling approach to galaxy clustering}",
      journal = {\jcap},
         year = 2023,
        month = apr,
       volume = {2023},
       number = {4},
          eid = {010},
        pages = {010},
          doi = {10.1088/1475-7516/2023/04/010},
archivePrefix = {arXiv},
       eprint = {2211.00660},
 primaryClass = {astro-ph.CO},
       adsurl = {https://ui.adsabs.harvard.edu/abs/2023JCAP...04..010H}
}
\bibliographystyle{aasjournalv7}

\end{document}